\documentstyle[12pt]{article}
\input epsf
\begin {document}
\def \a {\alpha}
\def \b {\beta}
\def \e {\epsilon}
\def \o {\omega}
\def \O {\Omega}
\def \d {\delta}

\def \g {\gamma}

\def \L {\Lambda}
\def \p {\pi}

\def \m {\mu}

\def \BE {\begin{equation}}
\def \EE {\end{equation}}
\def \BEA {\begin{eqnarray}}
\def \EEA {\end{eqnarray}}

\def \CB {\begin{centerline}}
\def \CE {\end{centerline}}
\def \bbox {\bf } 
\title{
Semiconductor Lasers and Kolmogorov Spectra.
}
\author{ Yuri V. Lvov$^{1,2}$ and Alan C. Newell$^{1,3}$}
\maketitle

$^1$ Department of Mathematics, The University of Arizona, Tucson AZ  
85721 USA 

$^2$ Department of Physics, The University of Arizona, Tucson AZ 85721 USA

$^3$ Mathematical Institute, The University of Warwick, Coventry CV47AL UK

\abstract{
In this letter, we make a prima facie case that there could be distinct 
 advantages to exploiting a new class of finite flux equilibrium 
solutions of the Quantum Boltzmann equation in semiconductor lasers.} 
\section{Introduction} At first sight, it may very well seem that the two 
subjects linked in the title have little in common. What do semiconductor 
lasers have to do with behavior normally associated with fully developed 
hydrodynamic turbulence?  In order to make the connection, we begin by 
reviewing the salient features of semiconductor lasers. In many ways, they 
are like two level lasers in that the coherent light output is associated 
with the in phase transitions of an electron from a higher to lower energy 
state. In semiconductors, the lower energy state is the valence band from 
which sea electrons are removed leaving behind 
positively charged holes. The higher energy state is the conduction band. 
The quantum of energy released corresponds to an excited electron in the 
conduction band combining  with a hole in the lower band below the 
bandgap.  Bandgaps, or forbidden energy zones are features of the energy 
spectrum of an electron in periodic potentials introduced  in this case  
by the periodic nature of the semiconductor lattice.

However, there are two  important ways in which the semiconductor laser 
differs from and is more complicated than the traditional two-level laser 
model.  First, there is a continuum of bandgaps parameterized by the 
electron momentum ${\bbox k}$ and the laser output  is a weighted sum of 
contributions from polarizations corresponding to electron-hole pairs  at 
each  momentum value. In this feature, the semiconductor laser resembles 
an inhomogeneosly broadened two level laser. Second, electrons and holes  
interact  with each other via Coulomb forces. Although this interaction is 
screened by the presence of many electrons and holes, it is nonetheless 
sufficiently strong to lead to a nonlinear coupling between electrons and 
holes at different momenta. The net effect of these collisions is a 
redistribution  of carriers (the common name for both electrons and holes) 
across the momentum spectrum. In fact it is the fastest ($\approx 100$ 
fs.) process (for electric field pulses of duration greater than 
picoseconds) and because of this, the gas of carriers essentially  relaxes 
to a distribution corresponding to an equilibrium of this collision 
process.  This equilibrium state is commonly taken to be that of 
thermodynamic equilibrium for fermion gases, the Fermi-Dirac 
distribution characterized by two parameters, the chemical potential 
$\mu$ and temperature $T$, slightly modified by the presence of broadband 
pumping and damping.

But the Fermi-Dirac distribution is not the only equilibrium of the 
collision process. There are other stationary solutions, called finite flux 
equilibria, for which there is a finite and constant flux of carriers and 
energy across a given spectral window. The Fermi-Dirac solution has zero 
flux of both quantities. It is the aim of this letter to suggest that 
these finite flux equilibria are more relevant to situations in which 
energy and carriers are added in one region of the spectrum, redistributed 
via collision processes to another region where they are absorbed. 
Moreover, it may be advantageous to pump the laser in this way because 
such a strategy may partially overcome the deleterious effects of Pauli 
blocking.  The Pauli exclusion principle means that two electrons with the 
same energy and spin cannot occupy the same state at a given momentum. 
This leads to inefficiency because the pumping is effectively multiplied 
by a factor $(1-n_s({\bbox k})), s=e,h$ for electrons and holes 
respectively, denoting the probability of not finding electron (hole) in a 
certain ${\bbox k}$ (used to denote both momentum and spin) state. But, 
near the momentum value corresponding to the lasing frequency $\o_L$, 
$n_s({\bbox k})$ is large ($n_e({\bbox k})+n_h({\bbox k})$ must exceed 
unity) and Pauli blocking significant. Therefore, pumping the laser in a 
window about $\o_0>\o_L$ in such a way that one balances the savings 
gained by lessening the Pauli blocking (because the carriers density 
$n_s({\bbox k})$ decreases with $k=|{\bbox k}|$) with the extra input 
energy required (because $k$ is larger), and then using the finite flux 
solution to transport carriers (and energy) back to lasing frequency, 
seems an option worth considering. The aim of this letter is to     
demonstrate, using the simplest possible model, that this alternative is 
viable. More detailed results using  more sophisticated (but far more 
complicated) models will be given later.

These finite flux equilibria are the analogies of the Kolmogorov spectra 
associated with fully developed, high Reynolds number hydrodynamic 
turbulence and the wave turbulence of surface gravity waves on the sea. In 
the former context, energy is essentially added at large scales (by 
stirring or some instability mechanism), is dissipated at small 
(Kolmogorov and smaller) scales of the order of less than the inverse 
three quarter power of the Reynolds number. It cascades via nonlinear 
interactions from the large scales to the small scales through a window of 
transparency (the inertial  range in which neither forcing nor damping is 
important) by the constant energy flux Kolmogorov solution. Indeed, for 
hydrodynamic turbulence,  the analogue to the Fermi-Dirac distribution,
the Rayleigh-Jeans spectrum of 
equipatitions, is irrelevant altogether.  The weak turbulence of surface 
gravity  waves is the classical analogue of the case of weakly interacting 
fermions. The mechanism for energy and carrier density (particle number) 
transfer is "energy" and "momentum" conserving binary collisions 
satisfying the "four wave resonance" conditions \BE {\bbox k}+{\bbox 
k_1}={\bbox k_2}+{\bbox k_3}, \ \ \ \ \o({\bbox k})+\o({\bbox 
k_1})=\o({\bbox k_2})+\o({\bbox k_3}).  \label{one} \EE In the 
semiconductor context, $\hbar \o({\bbox k})=\hbar \o_{\rm gap} 
    +\e_e({\bbox k})+\e_h({\bbox k})$ (which can be well approximated by 
    $\a+\b k^2$) where $\hbar \o_{\rm gap}=\e_{\rm gap}$ corresponds to 
the minimum bandgap and  $\e_e({\bbox k})$, $\e_h({\bbox k})$ are electron 
and hole energies. In each case, there is also a simple relation $E({\bbox 
k})=\o n({\bbox k})$ between the spectral energy density $E({\bbox k})$ 
and carrier (particle number) density $n({\bbox k})$. As a consequence of 
conservation of both energy and carriers, it can be argued (schematically 
shown in Figure 1 and described in its caption), that the flux energy (and 
some carriers) from intermediate momentum scales (around $k_0$ say) at 
which it is injected,  to higher momenta (where it is converted into heat) 
must be accompanied by the flux of carriers and some energy from $k_0$ to 
lower momenta at which it will be absorbed by the laser. It is the latter 
solution that we plan to exploit.

\section{Model}
We present the results of a numerical simulation of a greatly simplified 
model of semiconductor lasing in which we use parameter values which are 
realistic but make fairly severe approximations in which we (a) assume 
that the densities of electrons and holes are the same (even though their 
masses differ considerably) (b) ignore carrier recombination losses 
and (c) model the collision integral by a differential approximation
\cite{Optical}, \cite{HASS}, \cite{ZLF} 
 in 
which the principal contributions to wavevector quartets  satisfying 
(\ref{one}) are assumed to come from nearby neighbors. Despite the 
brutality of the approximations, the results we obtain are qualitatively 
similar to what we obtain using more sophisticated and 
complicated descriptions.

The semiconductor Maxwell-Bloch equations are \cite{Koch},\cite{Koch1},
\BEA \frac{\partial e}{\partial t}&=& i \frac{\O}{2\e_0}\int\mu_k p_k 
d{\bbox k} - \g_E e ,\label{two}\\
\frac{\partial p_k}{\partial t}&=&(i\O-i\o_k-\g_P)p_k - \frac{i 
\mu_k}{2\hbar}(2n_k-1)e,\label{three}\\
\frac{\partial n_k}{\partial t}&=&\L(1-n_k) -\g_k n_k +\left(
\frac{\partial n_k}{\partial t} \right)_{\rm collision} -\frac{i}{2\hbar}
\left( \mu_k p_k e^*-\m_k p_k^* e\right). \label{four}\EEA
Here $e(t)$ and $p_k(t)$ are the electric field and polarization at 
momentum $\bbox k$ envelopes of the carrier wave $\exp{(-i\O t + i K z)}$ 
where $\O$ is the cavity frequency (we assume single mode operation only) 
and $n(k)$ is the carrier density for electrons and holes. The constants 
$\g_E$, $\g_P$  model electric field and homogeneous broadening losses, 
$\e_0$ is dielectric constant, $\mu_k$ is the weighting accorded to 
different $\bbox k$ momentum (modeled by 
$\mu_k=\mu_{k=0}/(1+\e_k/\e_{\rm gap})$), $\L_k$ and $\g_k$ 
represent carrier pumping and damping. In (\ref{four}), the collision term 
is all important and is given by 
\BEA 
\frac{\partial} {\partial t}n_{k}=
4\pi \int |T_{k k_1 k_2 k_3}|^2    
\left( n_{\bbox k_2}n_{\bbox k_3}(1-n_{\bbox k_1}-n_{\bbox k})+ 
   n_{\bbox k}n_{\bbox k_1}(n_{\bbox k_2}+n_{\bbox 
k_3}-1)\right)\cr
 \times\d({\bbox k}+{\bbox k_1}-\bbox{k_2}-{\bbox k_3}) 
\d(\o_{k}+\o_{k_1}-\o_{k_2}-\o_{k_3})  d{\bbox k_1}d{\bbox 
k_2} d {\bbox k_3} ,  \label{five}\EEA 
where 
$T_{k k_1 k_2 k_3}$ is the coupling 
coefficient measuring mutual electron and hole interactions. We make the 
weak assumption that all fields are isotropic and make a convenient 
transformation from $k$ ($=|{\bbox k}|$) to $\o$ via the dispersion 
relation $\o=\o({\bbox k})$ defining the carrier density $N_\o$ by $\int 
N_\o d \o = \int n({\bbox k}) d{\bbox k}$ or $N_\o=4\p k^2 d k /(d\o)
n({\bbox k})$.  
Then, in the differential approximation, (\ref{five}) can be written as 
both, \BE   \frac{\partial N_\o}{\partial t} = \frac{\partial^2 
K}{\partial\o^2} \ \ \ \ \ \ {\rm and \ \ \ \ } \frac{\partial \o 
N_\o}{\partial t} = - \frac{\partial }{\partial\o}
 (K-\o\frac{\partial K}{\partial\o}), \label{six}\EE 
with
$$\\ \ \ \ \ K=-I\o^s \left( n^4_o\left(n_\o^{-1}\right)'' 
+n_\o^2 \left(\ln n_\o\right)''\right), \ \ \ ()'=\frac{\partial 
}{\partial\o}, \ \ \ n_\o=n({\bbox k}(\o)),$$ 
where $s$ is the number computed from the dispersion relation, the 
dependence of 
$T_{k k_1 k_2 k_3}$ on ${\bbox k}$ and dimensions ($s$ is of the order of 
 $7$ for semiconductors.) The conservation forms of the equations for 
$N_\o$ and $E_\o=\o N_\o$ allow us identify $Q=\frac{\partial K}{\partial 
\o}$ (positive if carriers flow from high to low momenta) and 
$P=K-\o\frac{\partial K}{\partial\o}$ (positive if energy flows from low 
to high momenta) as the fluxes of carriers and energy respectively.  
Moreover, the equilibrium solutions are now all transparent.  The general 
stationary solution of (\ref{six}) is the integral of $K=Q\o+P$ which 
contains four parameters, two (chemical potential and temperature) 
associated with the fact that $K$ is a second derivative, and two constant 
fluxes $Q$ and $P$ of carriers and energy.  The Fermi-Dirac solution 
$n_\o=(\exp{(A\o+B)}+1)^{-1}$, the solution of $K=0$, has zero flux. We 
will now solve (\ref{two}), (\ref{three}) and (\ref{four}) after angle 
averaging (\ref{four}) and replacing $4\pi k^2 \frac{\partial k}{\partial 
\o}\left(\frac{\partial n_k}{\partial \o}\right)_{\rm collision}$ by 
$\frac{\partial^2 K}{\partial\o^2}$. The value  of the constant $I$ is 
chosen to ensure that solutions of (\ref{six}) relax in a time of 100 fs.  

\section{Results}
We show the results in figures 2, 3, and 4. First, to test accuracy,
we show, in Figure 2, the relaxation of (\ref{six}) to a pure Fermi-Dirac 
spectrum in the window $\o_L=1<\o<\o_0=2$. The boundary conditions 
correspond to $P=Q=0$ at both ends.  We then modify boundary
conditions to read $Q=Q_0>0$  and $P=0$ at both ends.  Next, in Figure 3 and 4, we show the 
results of two experiments in which we compare the efficiencies  of two 
experiments in which we arrange to (i) pump broadly so that the effective 
carrier distribution equilibrium has zero flux and (ii) pump carriers and 
energy into a narrow band of frequencies about $\o_0$ and simulate this by 
specifying carrier and energy flux rates $Q_L$ and $P_L=-\o_L Q_L$ 
($P_L$ chosen so that the energy absorbed by the laser is consistent with 
the number of carriers absorbed there) at the boundary $\o=\o_0$. 
$\o=\o_L$ is the frequency at which the system lases.

In both cases, the rate of addition of carriers and energy is
(approximately) the same. The results support the idea that it is
worth exploring the exploitation of the finite flux equilibrium. The
carrier density of the equilibrium solutions at $\o_0$ is small thus
making pumping more efficient there. The output of the laser is
greater by a factor of 10.  While we do not claim that, when all
effects are taken account of, this advantage will necessary remain, we
do suggest that the strategy of using finite flux equilibrium
solutions of the Quantum Boltzmann equation is worth further
exploration.
\section{Acknowledgments} We are grateful for support from AFOSR 
Contract 94-1-0144-DEF and F49620-97-1-0002.
\section{Figure Captions.}
\begin{itemize}\item Figure 1 \\
Carriers and energy are added at $\o_0$ at rates $Q_0$ and $\o_0 Q_0$. 
Energy and some carriers are dissipated at $\o_R>\o_0$ (an idealization) 
and carriers and some energy are absorbed by the laser at $\o_L$. (The 
carriers number will build until the laser switches on.) A little 
calculation shows $Q_L=Q_0 (\o_R-\o_0)/(\o_R-\o_L)$,  
$Q_R=Q_0(\o_L-\o_0)/(\o_R-\o_L)$, $P_R=Q_0 \o_R (\o_0-\o_L)/(\o_R-\o_L)$, 
$P_L= \o_L Q_0 (\o_0-\o_R)/(\o_R-\o_L)$.  Finite flux stationary solutions 
are realized in the windows $(\o_L,\o_0)$ and $(\o_0,\o_R)$ although in 
practice there will be some losses through both these regions.  \item 
{Figure 2.}\\ To test accuracy we take some initial distribution function 
  (thin line) and plot its time evolution as described by (\ref{six}) with 
  boundary conditions $P=Q=0$ at both ends. The distribution function 
  relaxes to Fermi-Dirac state (thick line). Several intermediate states 
  is shown by long-dashed and short-dashed lines (Figure 2a). We then 
  modify boundary conditions to $P=0,Q=Q_0>0$ at both ends.  Then initial 
  distribution function (thick line) relaxes to finite-$Q$-equilibria as 
  shown by long-dashed line. We then change boundary conditions to $P=0, 
  Q=Q_1>Q_0$ at both ends, so that distribution function is shown by 
  short-dashed line.  Increasing $Q$ at boundaries even further, so that 
  $P=0, Q=Q_2>Q_1$ at both ends, the distribution function is given by 
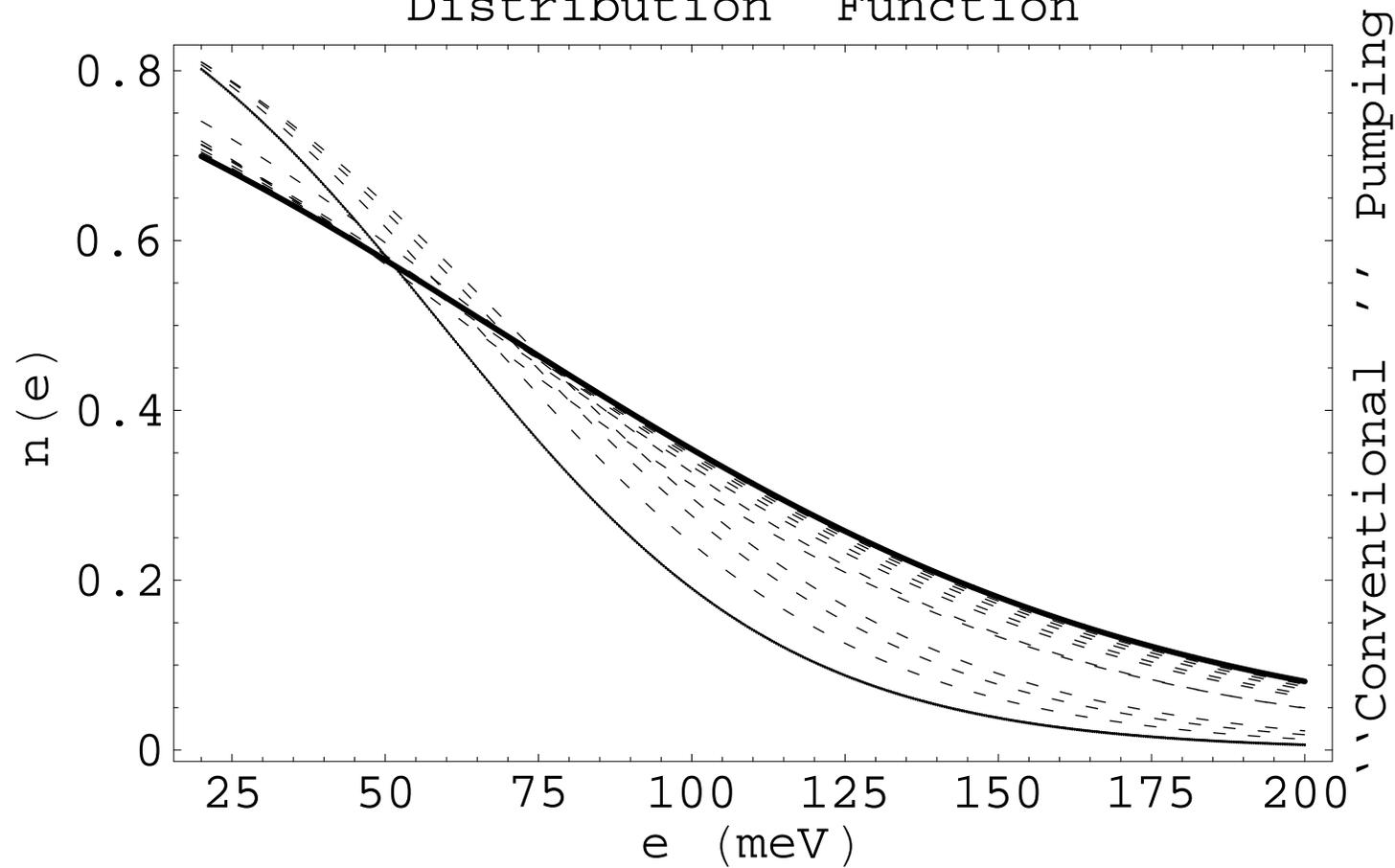
  dotted line (Figure 2b).  \item Figure 3.\\ We now solve 
(\ref{two}-\ref{four}) with the collision term given by (\ref{six}).  We 
pump broadly, so that the effective carrier distribution has zero flux. 
The initial distribution function (thin line) builds up because of a 
global pumping (dashed lines), until the laser switches on. The final 
(steady) distribution function is shown by thick solid line (Figure 3a).
The output power (in arbitrary units)  as a function of time (measured
in relaxation times $\simeq 100$fs) is also shown (Figure 3b).  \item 
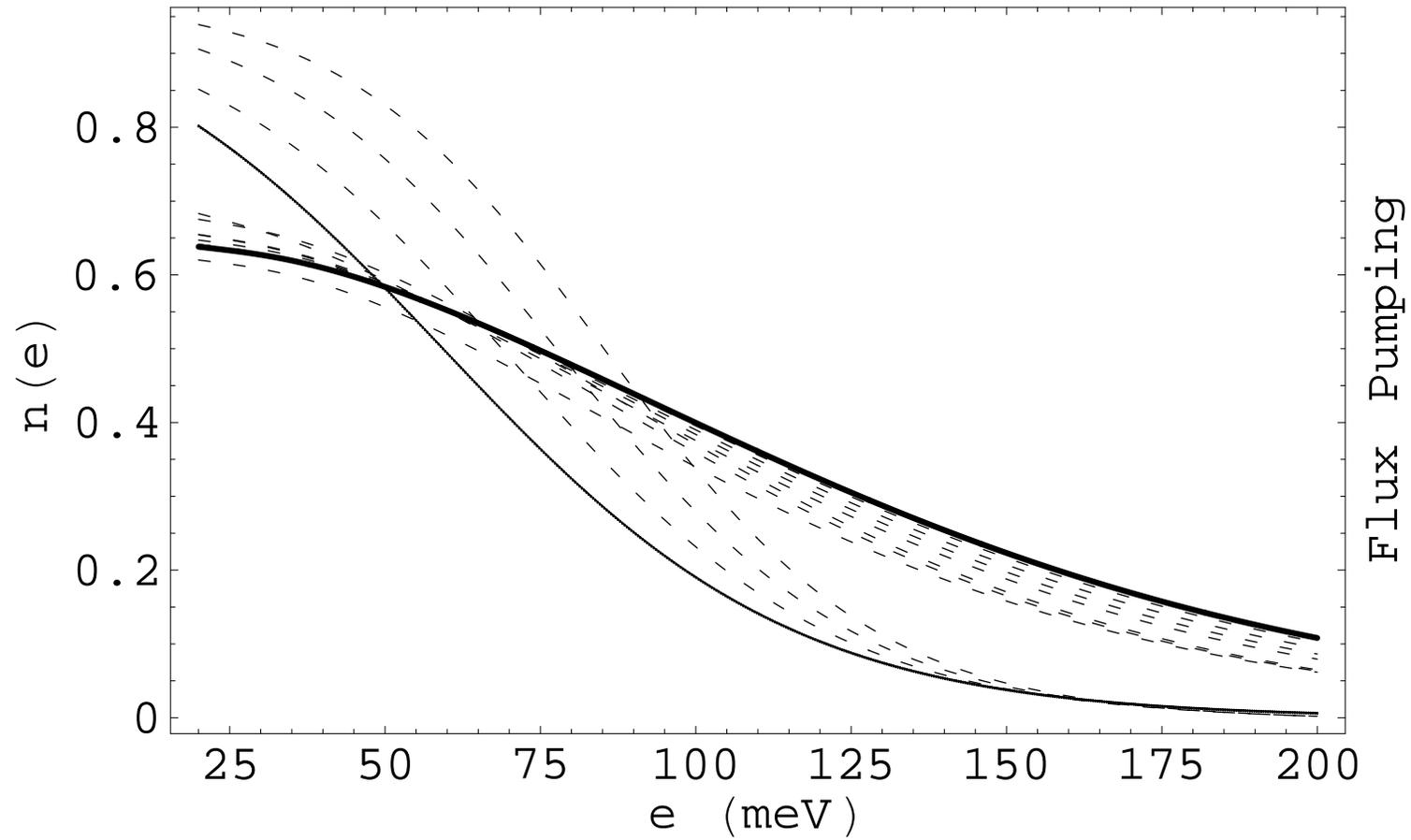
Figure 4.\\ We pump in the narrow region around $\o_0\simeq 200 {\rm meV}$ 
and we model this by specifying carrier and energy flux rates $Q_L$ and 
$P_L=-\o_L Q_L$.  The initial distribution function (thin line) builds up 
because of influx of particles and energy from right boundary (dashed 
  lines), until the laser switches on. The final (steady) distribution 
  function is shown by thick solid line and corresponds to a flux of 
  particles and energy from right boundary (where we add particles and 
  energy) to the left boundary, where the system lases (Figure 4a).
  The   output power as a function of time is also shown (Figure 4b).  
  \end{itemize} 
\newpage
\begin{centerline}{\LARGE Figure 1}\end{centerline}

\

\


\setlength{\unitlength}{.8cm}

\begin{picture}(16,4)(0,1)
\put(0,1.5){\line(1,0){16}}

\put(8,1.3){\line(0,1){.4}}
\put(8,1){\makebox(0,0){$\omega_0$}}

\put(8,2.2){\vector(0,1){1.6}}
\put(8,0.5){\makebox(0,0){IN}}
\put(7.6,4.3){$Q_0,\omega_0 Q_0$}

\put(1.25,1.3){\line(0,1){.4}}
\put(1.25,1){\makebox(0,0){$\omega_L$}}

\put(14.6,1.3){\line(0,1){.4}}
\put(14.6,1){\makebox(0,0){$\omega_R$}}

\put(7,1.2){\line(0,1){2.8}}
\put(9,1.2){\line(0,1){2.8}}
\put(2.5,1.2){\line(0,1){2.8}}

\put(1,2){\vector(0,-1){.8}}
\put(1,0.5){\makebox(0,0){OUT}}
\put(15,2){\vector(0,-1){.8}}
\put(15.2,0.5){\makebox(0,0){OUT}}

\put(10.2,2){\vector(-1,0){1}}
\put(10.5,1.9){$Q_R$}

\put(9.2,2.8){\vector(1,0){1}}
\put(10.5,2.7){$P_R$}

\put(6.8,2){\vector(-1,0){1}}
\put(5,1.9){$Q_L$}

\put(5.8,2.8){\vector(1,0){1}}
\put(5,2.7){$P_L$}
\noindent
\noindent
\noindent

\end{picture}

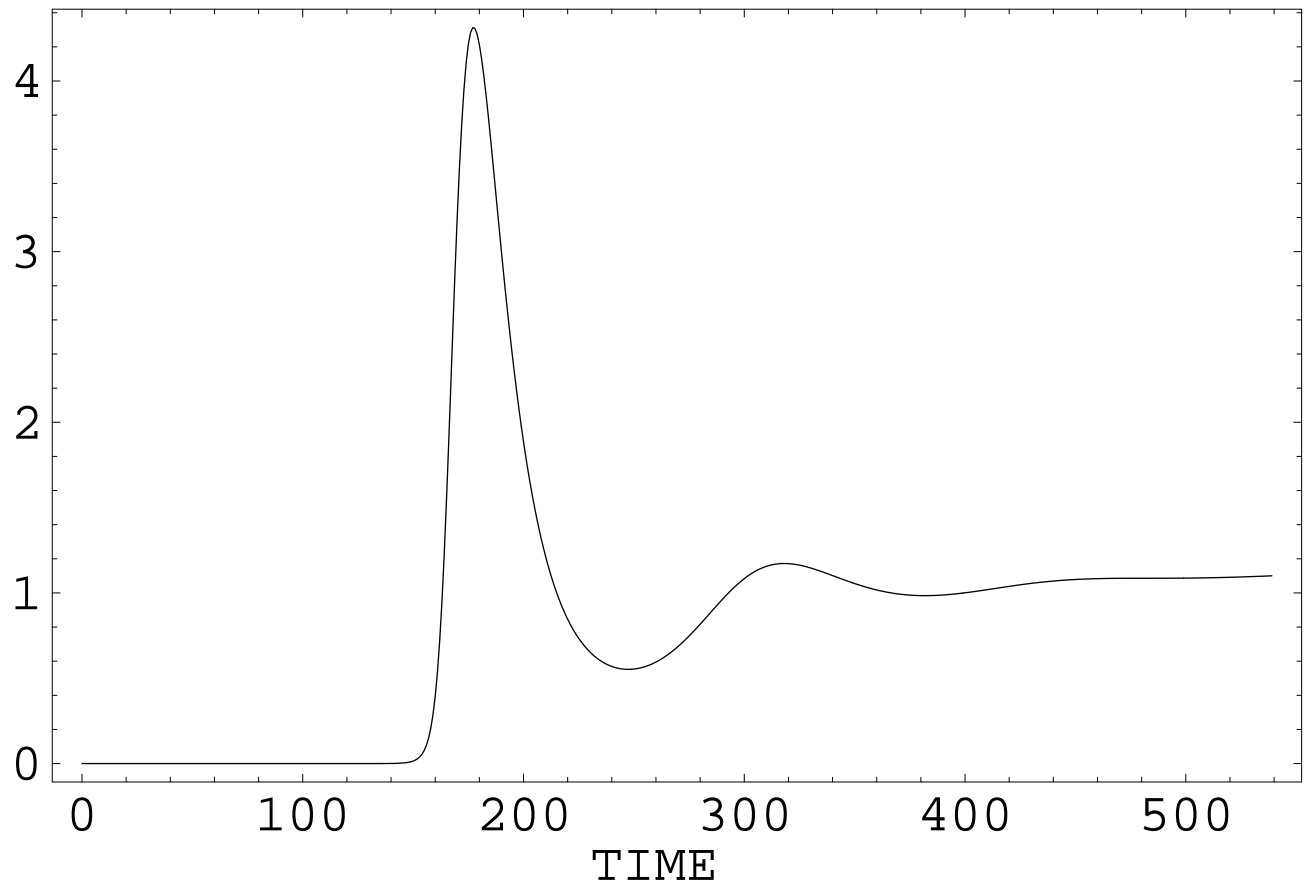
\end{document}